\documentclass[a4paper,12pt]{article}

\renewcommand{\baselinestretch}{1.2}

\usepackage{graphicx,url}
\usepackage[colorlinks=true,linkcolor=black,urlcolor=black]{hyperref}

\newcommand{\program}{SeCQC}
\newcommand{\version}{version~1.0}
\newcommand{\MainSite}{%
\url{http://www.quantumlah.org/publications/software/SeCQC/}}
\newcommand{\MainSiteSOMIM}{%
\url{http://www.quantumlah.org/publications/software/SOMIM/}}
\newcommand{\all}{%
\url{http://www.quantumlah.org/publications/software/SeCQC/all.tar.gz}}
\newcommand{\manual}{%
\url{http://www.quantumlah.org/publications/software/SeCQC/Manual.pdf}}
\newcommand{\source}{%
\url{http://www.quantumlah.org/publications/software/SeCQC/source.tar.gz}}
\newcommand{\executable}{%
\url{http://www.quantumlah.org/publications/software/SeCQC/secqc.tar.gz}}
\newcommand{\arXiv}{%
\url{http://arxiv.org/abs/0805.2847}}
\newcommand{\feedback}{%
\texttt{\href{mailto:secqc@quantumlah.org}{secqc@quantumlah.org}}}


\begin{document}\thispagestyle{empty}

\begin{center}\Large\textbf{SeCQC:}\\\large
An open-source program code for the numerical\\
\textbf{\emph{Se}}arch for the classical \textbf{\emph{C}}apacity of
\textbf{\emph{Q}}uantum \textbf{\emph{C}}hannels
\end{center}

\bigskip

\begin{center}
\ Jiangwei Shang,$^{1}$
\ Kean Loon Lee,$^{1,2}$
and Berthold-Georg Englert$^{1,3}$
\end{center}

\bigskip

\renewcommand{\baselinestretch}{1.0}\normalsize
\begin{center}\small\itshape
\begin{tabular}{r@{}p{0.72\textwidth}}
$^1$&
Centre for Quantum Technologies\newline
National University of Singapore\newline
3 Science Drive 2, Singapore 117543, Singapore\\[0.5ex]
$^2$&
Graduate School for Integrative Sciences and Engineering\newline
National University of Singapore\newline
28 Medical Drive, Singapore 117456, Singapore\\[0.5ex]
$^3$&
Department of Physics\newline
National University of Singapore\newline
2 Science Drive 3, Singapore 117542, Singapore\\[0.5ex]
\end{tabular}
\end{center}

\bigskip\bigskip\bigskip

\begin{quote}
\centerline{\Large\textbf{Abstract}}
\program\ is an open-source program code which implements a Numerical Search for
the classical Capacity of Quantum Channels (\program) by using an iterative method.
Given a quantum channel, \program\ finds the statistical operators and POVM outcomes that
maximize the accessible information, and thus determines the classical capacity of the quantum channel.
The optimization procedure is realized by using a steepest-ascent method that follows the
gradient in the POVM space, and also uses conjugate gradients for speed-up.
\end{quote}

\vfill
\centerline{This manual is for \version.}

\newpage
\tableofcontents

\section{License Agreement}
\program\ is an open-source program that, given a quantum channel,
implements a Numerical Search for the classical Capacity of Quantum Channels.
It is a derivative of the open-source program code SOMIM (see Ref.~\cite{SOMIMmanual}).
Copyright \copyright\ 2010 J.W. Shang, K.L. Lee and B.-G. Englert.

\program\ is a free software: You can redistribute it and/or modify it under
the terms of the GNU General Public License Version 3 as published by the Free
Software Foundation.

\program\ is distributed in the hope that it will be useful, but WITHOUT
ANY WARRANTY; without even the implied warranty of FITNESS or MERCHANTABILITY
FOR PARTICULAR PURPOSE.
See the GNU General Public License at
\url{http://www.gnu.org/licenses/} for details.

\section{What can \program\ be used for?}
\begin{figure}[htb]
\includegraphics[width=\textwidth]{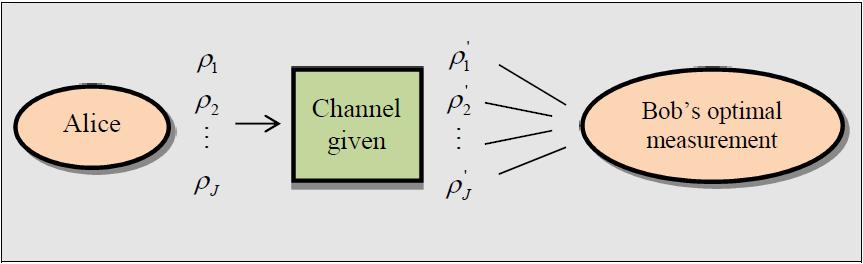}
\caption{\label{channel}Schematic setup of quantum communication scenario.}
\end{figure}
\textbf{Kraus representation of a Quantum Channel:}
We use a set of Kraus operators $K_m (m=1,...,N)$ to represent the quantum channel, such that
\begin{equation}
  \sum_{m=1}^N K_m^\dagger K_m=1\,.
\end{equation}
Consider the following quantum communication scenario shown in Fig.~\ref{channel}.
Alice sends a set of quantum states $\mathcal{E}=\{\rho_j\mid j=1,2,...,J\}$ with $\rho_j\geq 0$ to Bob through a quantum channel $M$, such that
\begin{equation}
  \rho = \sum_{j=1}^J\rho_j\quad
  \mbox{with}\;
  \mathrm{tr}\{\rho\}=1\,.
\end{equation}
The states after passing through the channel are given by
\begin{equation}
  \rho_j^\prime = \sum_{m=1}^N K_m \rho_j K_m^\dagger =M\rho_j\,.
\end{equation}
Bob performs a generalized measurement, specified by a positive-operator-valued measure (POVM),
on the state he receives.
The POVM with outcomes $\Pi_k (k=1,2,...,K)$ decomposes the identity,
\begin{equation}
  \sum_{k=1}^K\Pi_k=1\quad
  \mbox{with}\;
  \Pi_k\geq 0 \,.
\end{equation}
Then, the joint probability to receive the \emph{j}th state and get the \emph{k}th outcome is
\begin{equation}
  p_{jk}=\mathrm{tr}\{\rho_j^\prime\Pi_k\}\,,\quad
  \sum_{j,k}p_{jk}=1\,.
\end{equation}
Bob's figure of merit is the \emph{mutual information}
\begin{equation}
  I(\mathcal{E}; \Pi)=\sum_{j=1}^J\sum_{k=1}^K
  p_{jk}\log_2\frac{p_{jk}}{p_{j\cdot}p_{\cdot k}}\,.
\end{equation}
where $p_{j\cdot}$ and $p_{\cdot k}$ are the marginal probabilities,
\begin{equation}
  p_{j\cdot}=\sum_{k}p_{jk}=\mathrm{tr}\{\rho_j\}\,,\quad
  p_{\cdot k}=\sum_{j}p_{jk}=\sum_{m}\mathrm{tr}\{\rho (K_m^\dagger \Pi_kK_m) \}\,.
\end{equation}
As stated, the $\rho_j$s are normalized such that their traces equal the probabilities of receiving them.

\textbf{Classical Capacity of Quantum Channels:}
Generally, the classical capacity of a quantum channel can be defined as
the maximum accessible information with respect to both statistical operators and POVM outcomes,
\begin{equation}
  \mathrm{\emph{C}}=\max_{\mathcal{E}}\max_{\Pi}I(\mathcal{E};\Pi)\,.
\end{equation}
Given a certain quantum channel, \program\ finds the statistical operators as well as POVM outcomes
that maximize the accessible information (AI), and thus determines the
classical capacity of the quantum channel.

The calculation is performed using a combination of the steepest-ascent method (see Ref.~\cite{rehacek2005} and Section 11.5 in Ref.~\cite{review2007}) and the conjugate-gradients (CG) method~\cite{NRC}.
The percentage chance to calculate with one method or the other can be specified by the user
(see Section 4 below).
The implementation in \program\ also makes use of the golden-section search method.

\section{Download and Compile}
The complete set of files, including this manual, are available at the
\program\ site: \MainSite.
Download \all, if you want to have the complete collection of files.
Just this manual is fetched from \manual.
The Windows executable file for \program\ can be downloaded from \executable.
If you intend to modify the code, you can download the source files from
\source.

The program is written in C++ and the graphic user interface (GUI) is
implemented using wxWidgets (\url{http://www.wxwidgets.org/}).
Here are the instructions for compiling \program:
\begin{enumerate}
\item Install wxWidgets from \url{http://www.wxwidgets.org/downloads/}.
\item If you are working in Windows, you need to install MinGW
  (\url{http://www.mingw.org/download.shtml}) and MYSY
  (\url{http://www.mingw.org/msys.shtml}) as well.
\item When wxWidgets and MinGW are configured, you can compile
  \program\ by executing \textbf{g++ MI.cpp `wx-config --libs` `wx-config
          --cxxflags` -o YourProgramName} in MSYS shell.
\item If you face problems running the program in a Linux environment,
  try \textbf{export LD\_LIBRARY\_PATH=/usr/local/lib}.
\item The executable file is compiled under Windows XP Service Pack 3,
  with wxWidgets 2.8.10, MinGW 5.1.6 and MSYS 1.0.11.
\end{enumerate}

\section{How to use the program}
\begin{figure}[p!]\centering
\includegraphics[width=\textwidth]{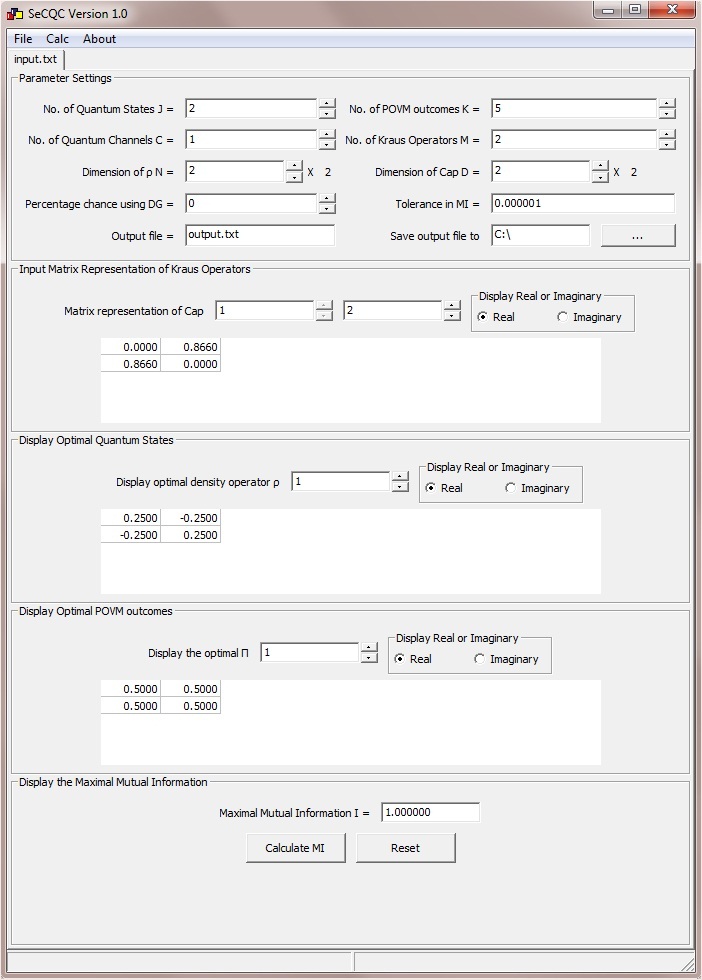}
\caption{\label{GUI}The graphical user interface (GUI) of the program.}
\end{figure}
The GUI of \program\ is shown in Fig.~\ref{GUI}.
In the first box labeled as ``Parameter Settings'', $J$ is the number of
statistical operators $\rho_j$.
The current maximum possible value is $J=30$.
Parameter $K$ is the initial number of POVM outcomes, with the largest
possible value being $K=30$.
The third and fourth fields are the number $C$ of Quantum Channels to be inserted and the number $M$ of
Kraus operators for each channel respectively.
The next two fields are the dimension $D$ of Kraus operators and the dimension $N$ of statistical operators respectively
with 30 being their highest possible value.
All the maximum values mentioned above can be changed by modifying the source code.
The seventh field is the percentage chance to use the steepest-ascent
method to perform maximization in an iteration;
this parameter controls the relative frequency of using the direct
or the conjugate gradient.
The eighth field gives the tolerance in the accessible information, the stopping
criteria for the computation;
the calculation stops when the difference in accessible information between
the current iteration and the previous iteration is less than half of the sum
multiplied by the tolerance plus the $\mathrm{machine\_epsilon}$ $\epsilon_m$
(also termed as the machine accuracy, typical value for
double precision is around $1.6\times 10^{-16}$),
i.e. when $2.0\times(\mathrm{current} - \mathrm{previous}) \leq
\mathrm{tolerance}\times(\mathrm{current}+\mathrm{previous})+\epsilon_m$.
The ninth field is the name of the output file.
By default, the output file will be located at hard disk C. You can change
the output directory by clicking the ellipsis button ``...'' and choose your preferred location.

The next three boxes display the input Kraus operators $\{K_m\}_{m=1,\dots,M}$,
the optimal statistical operators $\{\rho_j\}_{j=1,\dots,J}$
and the calculated optimal POVM outcomes $\{\Pi_k\}_{k=1,\dots,K}$.
The spin buttons are used to switch between the various $K_{m}$s/$\rho_j$s/$\Pi_k$s,
while the small box beside the spin button is used to
choose to display the real or imaginary part of the chosen $K_{m}$s/$\rho_j$s/$\Pi_k$s.

The maximum accessible information for the given channel will be
displayed in the last box after the ``Calculate MI'' button is pressed.
All values will be reset to default when the ``Reset'' button is pressed.

\textbf{Important note:} The matrices for the $K_m$s must have the correct
dimension; they must satisfy the condition, i.e.
$\sum_{m}K_m^\dagger K_m=1$.

\section{How to import data}\label{importdata}
\begin{figure}[htbp!]\centering
\includegraphics[width=\textwidth]{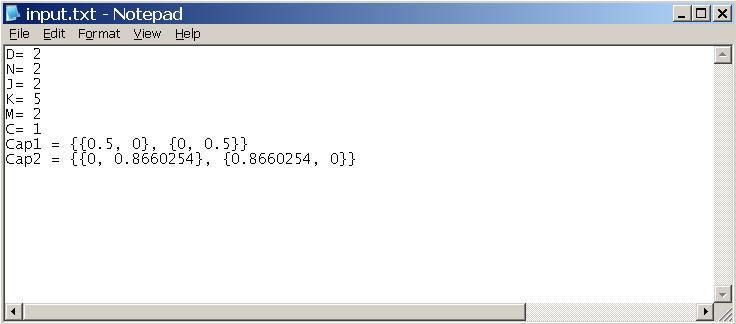}
\caption{\label{import}Example of an import file.}
\end{figure}

Data can be imported into \program\ using a text file that is possibly
generated by another program.
An example is shown in Fig.~\ref{import}.
When importing, the numbers after the equal signs will be read into the program.
The first line is the dimension $D$ of the Kraus operators.
The second line gives the dimension $N$ of the statistical operators.
The third line gives the number $J$ of the statistical operators and
the fourth line gives the number of outcomes $K$ of POVMs that the program should start calculating with.
And the last two lines give the number $M$ of the Kraus operators for each channel
and the number $C$ of Quantum Channels respectively.

The subsequent lines give the input matrices for the Kraus operators.
Each line will give only one operator.
For an operator represented in matrix form as
\begin{equation}
\left(\begin{array}{cc}
		0.1 & 0.3+0.5i \\
		0.3-0.5i & 0.6
		\end{array}\right),
\end{equation}
the input data should be formatted as
$\{\{0.1,0.3+0.5\mathrm{\texttt{I}}\},\{0.3-0.5\mathrm{\texttt{I}},0.6\}\}$.

Complex numbers are entered as
$\mathrm{RealPart}+\mathrm{ImaginaryPart}\,\mathrm{\texttt{I}}$, as illustrated by
$-3.1-4.5\mathrm{\texttt{I}}$.
Please note that the complex unit $i$ must be entered in upper case \texttt{I}
and it must be at the end of the entry.

\section{Meaning of output data}
\begin{figure}[p!]\centering
\includegraphics[width=\textwidth]{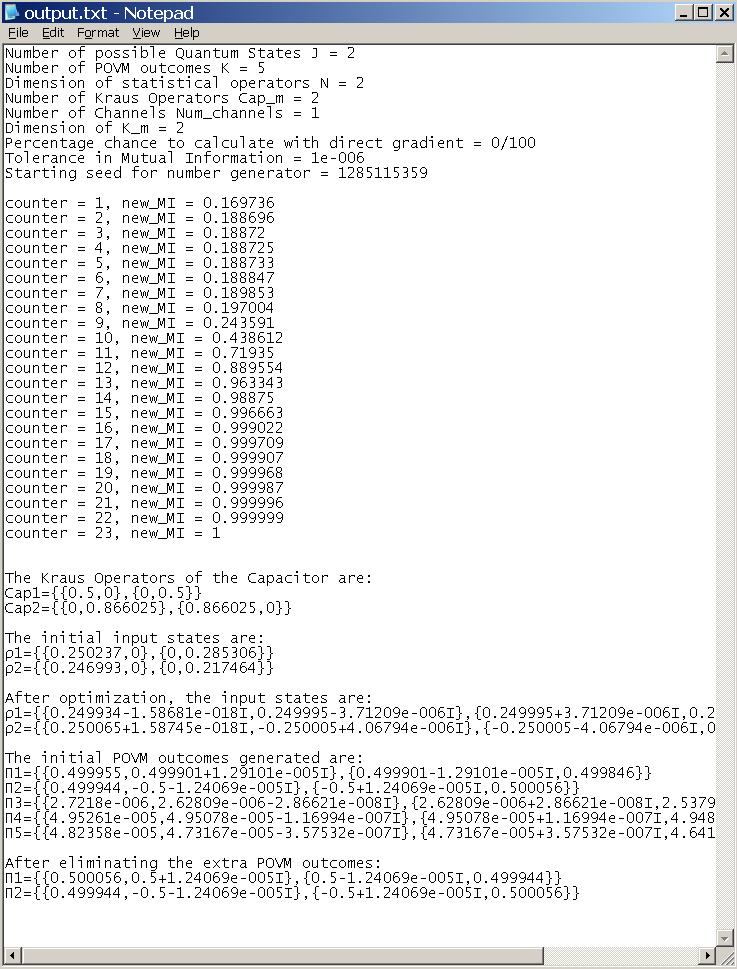}
\caption{\label{output}Example of output file.}
\end{figure}

A typical output file looks like Fig.~\ref{output}.
The first eight lines give the following information:
the number $J$ of statistical operators,
the initial number $K$ of POVM outcomes,
the dimension $N$ of the statistical operators,
the number $M$ of Kraus operators for each channel,
the number $C$ of Quantum Channels inserted,
the dimension $D$ of the Kraus operators,
the tolerance in the calculated mutual information,
and the seed for the random number generator.

The next block of lines gives the mutual information at the end of each
iteration. In the example shown in Fig.~\ref{output}, altogether 26 iterations have been performed
with the final accessible information being $\mathrm{AI}=1.000$ exactly.

The subsequent five blocks of lines give the $M$ Kraus operators for each channel,
the initial as well as the optimized $J$ statistical operators
and the $K$ outcomes of the optimal POVMs that correspond to the accessible information
calculated in the final round of iteration.
Each Kraus oeprator/statistical operator/POVM outcome is given in a single line in matrix
form, as explained in Section \ref{importdata}.

Among the $K$ outcomes, if any two outcomes, say $\Pi_{k_1}$ and
$\Pi_{k_2}$, give equivalent probabilities, i.e.
$p_{jk_1}p_{\cdot k_2}=p_{\cdot k_1}p_{jk_2}$ for all $j$, then these two
POVM outcomes are replaced with one new POVM outcome,
$\Pi_{k_1}+\Pi_{k_2}$, such that the new optimal POVM contains only
$K-1$ outcomes.
The last block of data in the output file gives the POVM after this
elimination process, i.e. the POVM is the optimal POVM with the least number
of outcomes.

\textbf{Caution:} As is the case for all steepest-ascent methods, there is the possibility of convergence towards a local, rather than a global, maximum. There is no absolute protection against this danger, but in practice one can fight it efficiently by running the program many times for comparison, with different seeds.
It also helps to start with a rather large $K$ value.

\section{Contact information}
Please send your comments, suggestions, or bug reports to the following email
account: \feedback

\section{Acknowledgments}
We acknowledge many valuable discussions with S.Y. Looi.
Centre for Quantum Technologies (CQT) is a Research Centre of Excellence funded by
Ministry of Education and National Research Foundation of Singapore.

\end{document}